\documentclass[twoside,leqno,twocolumn]{article}
\usepackage{ltexpprt}

\usepackage[utf8]{inputenc}
\usepackage[T1]{fontenc}

\usepackage{amsmath}
\usepackage{amssymb}
\usepackage{graphicx}
\usepackage{textgreek}
\usepackage{microtype}

\usepackage{url}

\newcommand{\set}[1]{\ensuremath{\{ #1 \}}}
\newcommand{\abs}[1]{\ensuremath{\lvert #1 \rvert}}
\newcommand{\Oh}[1]{\ensuremath{\mathsf{O}\!\left( #1 \right)}}
\newcommand{\oh}[1]{\ensuremath{\mathsf{o}\!\left( #1 \right)}}

\newcommand{\baseA}{\mathtt{A}}
\newcommand{\baseC}{\mathtt{C}}
\newcommand{\baseG}{\mathtt{G}}
\newcommand{\baseT}{\mathtt{T}}
\newcommand{\baseN}{\mathtt{N}}
\newcommand{\dnacomp}[1]{\ensuremath{\overline{#1}}}
\newcommand{\revcomp}[1]{\ensuremath{\overleftarrow{#1}}}

\newcommand{\rank}{\ensuremath{\mathsf{rank}}}
\newcommand{\select}{\ensuremath{\mathsf{select}}}
\newcommand{\LF}{\ensuremath{\mathsf{LF}}}
\newcommand{\find}{\ensuremath{\mathsf{find}}}
\newcommand{\locate}{\ensuremath{\mathsf{locate}}}
\newcommand{\parent}{\ensuremath{\mathsf{parent}}}
\newcommand{\countq}{\ensuremath{\mathsf{count}}}

\newcommand{\gindegree}{\ensuremath{\mathsf{in}}}
\newcommand{\goutdegree}{\ensuremath{\mathsf{out}}}
\newcommand{\glabel}{\ensuremath{\mathsf{label}}}
\newcommand{\gpred}{\ensuremath{\mathsf{pred}}}
\newcommand{\gkey}{\ensuremath{\mathsf{key}}}
\newcommand{\gvalue}{\ensuremath{\mathsf{value}}}
\newcommand{\gnode}{\ensuremath{\mathsf{node}}}
\newcommand{\gext}{\ensuremath{\mathsf{ext}}}

\newcommand{\kmer}[1]{$#1$\nobreakdash-mer}
\newcommand{\kcollection}[1]{$#1$\nobreakdash-collection}
\newcommand{\kequivalent}[1]{$#1$\nobreakdash-equivalent}
\newcommand{\orderk}[1]{order\nobreakdash-$#1$}
\newcommand{\LFmapping}{LF\nobreakdash-mapping}
\newcommand{\FMindex}{FM\nobreakdash-index}
\newcommand{\patternset}{\ensuremath{(\Sigma \setminus \set{\#, \$})^{\ast}}}

\newcommand{\SA}{\ensuremath{\mathsf{SA}}}
\newcommand{\BWT}{\ensuremath{\mathsf{BWT}}}
\newcommand{\Carray}{\ensuremath{\mathsf{C}}}
\newcommand{\LCP}{\ensuremath{\mathsf{LCP}}}
\newcommand{\bvIN}{\ensuremath{\mathsf{IN}}}
\newcommand{\bvOUT}{\ensuremath{\mathsf{OUT}}}

\title{Indexing Variation Graphs\thanks{Supported by the Wellcome Trust grant 098051.}}

\author{Jouni Sirén\thanks{Wellcome Trust Sanger Institute, UK. \texttt{jouni.siren@iki.fi}.}}

\date{}

\begin{document}

\maketitle

\begin{abstract}\small\baselineskip=9pt
Variation graphs, which represent genetic variation within a population, are replacing sequences as reference genomes. Path indexes are one of the most important tools for working with variation graphs. They generalize text indexes to graphs, allowing one to find the paths matching the query string. We propose using de Bruijn graphs as path indexes, compressing them by merging redundant subgraphs, and encoding them with the Burrows-Wheeler transform. The resulting fast, space-efficient, and versatile index is used in the variation graph toolkit vg.
\end{abstract}

\section{Introduction}

Sequence analysis pipelines typically start with mapping the reads from the sequenced genome to a \emph{reference genome} of the same species. As reference genomes are usually assembled from the genomes of a small number of individuals, they are biased towards those individuals. This \emph{reference bias} may affect the results of subsequent analysis, especially when the sequenced individuals are from different populations than the ones behind the reference genome.

\emph{Variation graphs} (also graph genomes, genome graphs, graph references, or reference graphs), which encode the genetic variation within a population as a graph, have been proposed as a solution to the reference bias \cite{Paten2014,Marcus2014,Church2015,Dilthey2015,Marschall2016}. These graphs are expected to replace sequences as reference genomes. The shift to graphs will likely take years, as new methods and tools are needed to replace those based on linear references.

The \emph{variation graph toolkit vg} \cite{Garrison2014-2016} is a community effort to develop such tools. This paper describes \emph{GCSA2}, the path index used in vg. A \emph{path index} is a generalization of text indexes for labeled graphs. Given a query string, the index finds the paths with a label matching the query. Indexing graphs is inherently hard, as the number of paths increases exponentially with path length. The design of a path index is hence a trade-off between maximum query length, index size, query performance, and pruning complex regions of the graph.

Mapping reads to a graph was first investigated by Schneeberger et~al.~\cite{Schneeberger2009}. Sirén et~al.\ developed \emph{GCSA}~\cite{Siren2014}, which generalized the FM\nobreakdash-index \cite{Ferragina2005a} (a text index based on the Burrows-Wheeler transform \cite{Burrows1994}) to directed acyclic graphs. GCSA depends on pruning the complex regions, as there are no limits on query length. Kim et~al.~\cite{Kim2015-2016} combined GCSA with the HISAT read aligner \cite{Kim2015}. The result was the first practical graph-based read aligner, though it uses the graph for more accurate mapping to a linear reference.

\emph{De~Bruijn graphs} can be used as \kmer{k} indexes of other graphs. There are already read aligners based on them \cite{Limasset2015,Liu2016}. The \emph{succinct de~Bruijn graph} of Bowe et.~al.~\cite{Bowe2012} encodes the graph with a generalization of the FM\nobreakdash-index. Rødland \cite{Roedland2013} proposed another similar generalization. Succinct de~Bruijn graphs can simulate \orderk{k} de~Bruijn graphs for multiple values of $k$ \cite{Boucher2014}, but they still need to store the graph explicitly for the largest value of $k$. Other representations include compacted \cite{Cazaux2014} and compressed \cite{Marcus2014} de~Bruijn graphs, which represent unary paths in the graph as single nodes or edges. Probabilistic de~Bruijn graphs \cite{Pell2012} use Bloom filters to support faster queries, at the expense of producing false positives. All these space-efficient representations require several bits per \kmer{k} for the graph, and more for mapping back to the indexed graph.

Some path indexes store the graph as a collection of sequences. BWBBLE \cite{Huang2013} uses the powerset alphabet for encoding substitutions and creates new sequences with a sufficient amount of context for other variants. vBWT \cite{Maciuca2016} encodes variant sites explicitly in the sequence as $X(A|B|C)Y$, using distinct separator symbols for each site. Queries in both BWBBLE and vBWT are slower than in ordinary FM\nobreakdash-indexes, as variant sites force the search to branch. The hypertext index \cite{Thachuk2013} works with graphs that have string labels on the nodes. The strings are indexed using an FM-index. Partial matches in the strings are combined into full matches with range queries in the edge matrix. While matches crossing one edge are easy to find, the approach becomes impractical with matches crossing multiple edges.

There are also structures (e.g.~\cite{Huang2010,Wandelt2013,Danek2014,Na2015,Na2016}) using graphs as a space-efficient way of indexing similar sequences. While the problem is different from indexing the paths in a graph, the techniques used are similar.

The above methods can be classified in three categories based on the data models they use. Many deal with graphs arising from \emph{aligned sequences}, assuming a shared global sequence with local variation. GCSA-based methods can index directed acyclic graphs, as well as cyclic graphs that are sufficiently similar to \emph{de~Bruijn graphs}. GCSA2, the hypertext index, and indexes based on de~Bruijn graphs work with \emph{arbitrary graphs}.

GCSA2 combines ideas from the original GCSA and from succinct de~Bruijn graphs. Conceptually it uses a de~Bruijn graph as a \kmer{k} index of a variation graph. The de~Bruijn graph is \emph{pruned} (compressed structurally) by using strings shorter than $k$ characters as nodes, if the shorter strings identify the start nodes of the corresponding paths uniquely. The pruned graph is encoded with a generalization of the FM\nobreakdash-index. GCSA2 often uses less space (e.g.~less than 1~bit per \kmer{k}) than other de~Bruijn graph-based indexes, which have to store some information for each \kmer{k} explicitly. The index also includes extensions based on suffix trees. The extensions are used for e.g.~finding maximal exact matches in the vg read aligner.

The main differences to the original GCSA are:
\begin{enumerate}

\item The graph encoding in GCSA2 has been optimized for small alphabets, improving query performance by up to an order of magnitude.

\item The construction algorithm stores the graphs on disk, reducing the memory requirements of building a whole-genome human index from hundreds of gigabytes to tens of gigabytes.

\item GCSA2 can index denser graphs, including cyclic graphs, by limiting maximum query length.

\item GCSA2 extends the functionality of the FM-index with suffix tree operations.

\end{enumerate}
The first two improvements can also be used with the original GCSA. The third point represents a different approach to indexing graphs. It was not possible the with the original GCSA, as the construction algorithm required a prefix-range-sorted graph. The extended functionality depends on the new construction algorithm and on limiting maximum query length.

This paper describes the GCSA2 data structure. Its uses in the vg toolkit will be discussed in the vg paper.

\section{Background}

\subsection{Strings}\label{sect:strings}

A \emph{string} $S[0, n-1] = s_{0} \dotsm s_{n-1}$ of length $\abs{S} = n$ is a sequence of \emph{characters} over an \emph{alphabet} $\Sigma = \set{0, \dotsc, \sigma - 1}$. \emph{Text} strings $T[0, n-1]$ are terminated by an \emph{endmarker} $T[n-1] = \$ = 0$ not found anywhere else in the text. A \emph{substring} of string $S$ is a sequence of the form $S[i, j] = s_{i} \dotsm s_{j}$. We call substrings of the type $S[0, j]$ and $S[i, n-1]$ \emph{prefixes} and \emph{suffixes}, respectively, and refer to substrings of length $k$ as \kmer{k}s. Substring $S[i, j]$ is a \emph{proper} substring of string $S$, if $S \ne S[i, j]$. We say that string $S'$ is a substring of string collection $\mathcal{S}$, if it is a substring of a string $S \in \mathcal{S}$.

Sometimes we consider \emph{infinite} character sequences $S = (s_{i})_{i \in Z}$, where set $Z$ is a contiguous infinite subset of $\mathbb{Z}$. The notion of substring generalizes to infinite sequences in a natural way. A substring of an infinite sequence $S$ is \emph{left-infinite} if it extends infinitely to the left, and \emph{right-infinite} if it extends infinitely to the right. A substring of a finite or infinite sequence $S$ is \emph{left-maximal} if it is left-infinite or a prefix, and \emph{right-maximal} if it is right-infinite or a suffix.

We are primarily interested in sequences over the \emph{DNA} alphabet $\set{\$, \baseA, \baseC, \baseG, \baseT, \baseN}$. Characters $\baseA$, $\baseC$, $\baseG$, and $\baseT$ are called \emph{bases}, while character $\baseN$ represents an arbitrary or unknown base. The alphabet may contain other technical characters in addition to the endmarker $\$$. Each character $c$ of the DNA alphabet has a \emph{complement} $\dnacomp{c}$ defined as $\dnacomp{\baseA} = \baseT$, $\dnacomp{\baseC} = \baseG$, $\dnacomp{\baseG} = \baseC$, $\dnacomp{\baseT} = \baseA$, and $\dnacomp{c} = c$ for other characters $c$. Given a DNA sequence $S$, its \emph{reverse complement} is the sequence $\revcomp{S}$ obtained by reversing the non-technical parts of the sequence and replacing each character with its complement. For example, $\revcomp{\mathtt{GATTACA}\$} = \mathtt{TGTAATC}\$$.

Given a string $S[0, n-1]$, we define $S.\rank(i, c)$ as the number of occurrences of character $c$ in the prefix $S[0, i-1]$. We also define
$$
S.\select(i, c) = \max \set{ j \le n \mid S.\rank(j, c) < i }
$$
as the position of the occurrence of character $c$ with rank $i > 0$.\footnote{These definitions are used in the SDSL library \cite{Gog2014b}. We assume for convenience that $S.\select(0, c) = -1$.} A \emph{bitvector} is a binary sequence supporting efficient $\rank$/$\select$ queries. \emph{Wavelet trees} \cite{Grossi2003} are space-efficient data structures that use bitvectors to support $\rank$/$\select$ queries on arbitrary strings.

Let $S$ be a string and $S'$ be a string or an infinite character sequence over alphabet $\Sigma$. We say that sequences $S$ and $S'$ \emph{prefix-match}, if $S$ is a prefix of $S'$ or $S'$ is a prefix of $S$. Set $\mathcal{S}$ of strings is \emph{prefix-free}, if no two strings $S, S' \in \mathcal{S}$ (with $S \ne S'$) prefix-match.

\subsection{Text Indexes}

The \emph{suffix tree} \cite{Weiner1973} is the most fundamental full-text index supporting substring queries. It is formed by taking the suffixes of the text, storing them in a trie, and compacting unary paths in the trie into single edges. Although fast and versatile, suffix trees are impractical with large texts, as they require much more space than the text itself.

\begin{figure}
\includegraphics{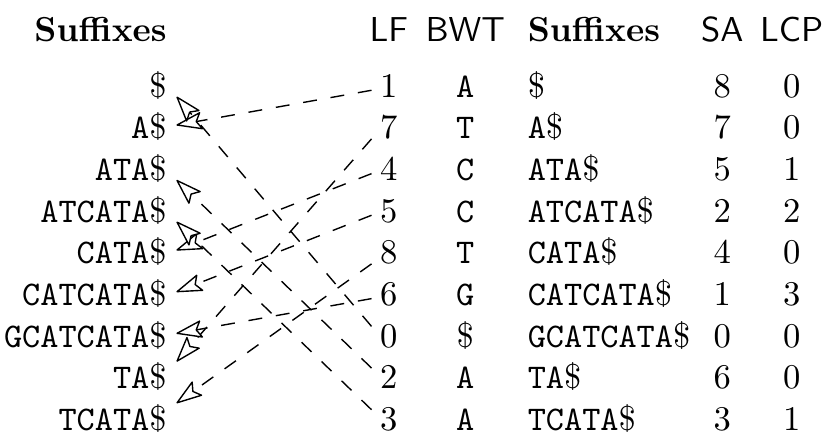}
\caption{\protect\LFmapping, BWT, suffix array, and LCP array for text $\mathtt{GCATCATA}\$$.}\label{figure:indexes}
\end{figure}

\emph{Suffix arrays} (SA) \cite{Manber1993} were introduced as a space-efficient alternative to the suffix tree. The suffix array of text $T[0, n-1]$ is an array of pointers $\SA[0, n-1]$ to the suffixes of the text in \emph{lexicographic order}. We find the occurrences of \emph{pattern} string $X$ in the text in $\Oh{\abs{X} \log n}$ time by using \emph{binary search} in the suffix array. The suffix array requires $n \log n$ bits of space in addition to the text, while its functionality is more limited than that of the suffix tree. See Figure~\ref{figure:indexes} for an example of the suffix array and related structures.

The \emph{Burrows-Wheeler transform (BWT)} \cite{Burrows1994} of text $T[0, n-1]$ is a permutation $\BWT[0, n-1]$ such that $\BWT[i] = T[(\SA[i]-1) \bmod n]$. Given the \emph{lexicographic rank} $i$ of suffix $T[\SA[i], n-1]$, we can use \emph{\LFmapping} on the BWT to find the lexicographic rank of the previous suffix $T[(\SA[i]-1) \bmod n, n-1]$. Let
$$
\LF(i) = \Carray[\BWT[i]] + \BWT.\rank(i, \BWT[i]),
$$
where $\Carray[c]$ is the number of occurrences of characters $c' < c$ in the BWT. Then $\SA[\LF(i)] = (\SA[i]-1) \bmod n$. We generalize the definition to any character $c \in \Sigma$:
$$
\LF(i, c) = \Carray[c] + \BWT.\rank(i, c).
$$
Let $X$ be a string. If there are $i$ suffixes $S'$ of text $T$ such that $S' < X$ in lexicographic order, then there are $\LF(i, c)$ suffixes $S'$ such that $S' < cX$.

We can use the BWT as a space-efficient text index. The \emph{\FMindex} \cite{Ferragina2005a} combines a representation of the BWT supporting $\rank$/$\select$ queries, the $\Carray$ array, and a set of \emph{sampled pointers} from the suffix array. It uses \emph{backward searching} to find the \emph{lexicographic range} of suffixes matching pattern $X$ (having $X$ as a prefix). If the lexicographic range matching suffix $X[i+1, \abs{X}-1]$ of the pattern is $\SA[sp, ep]$, the range matching suffix $X[i, \abs{X}-1]$ is $\SA[\LF(sp, X[i]), \LF(ep+1, X[i]) - 1]$. Matching the entire pattern takes $\Oh{\abs{X}}$ $\rank$ queries.

We use the sampled suffix array pointers for finding the text positions containing the occurrences. If $\SA[i]$ is not sampled, we iterate $\LF(i)$ until we find a sampled pointer. If we find a sample at $\SA[\LF^{k}(i)]$, we know that
$$
\SA[i] = (\SA[\LF^{k}(i)] + k) \bmod n.
$$
If we have sampled one out of $d$ suffix array pointers at regular intervals, finding each occurrence takes $\Oh{d}$ rank queries. If we also sample one out of $d'$ \emph{inverse suffix array} pointers\footnote{The suffix array is a permutation of $\set{0, \dotsc, n-1}$, and the inverse suffix array is the inverse permutation.}, we can \emph{extract} an arbitrary substring $X$ of the text using $\Oh{\abs{X}+d'}$ rank queries.

The \emph{longest-common-prefix array} (LCP array) \cite{Manber1993} is an integer array $\LCP[0, n-1]$, where each value $\LCP[i]$ tells the length of the longest common prefix of suffixes $T[\SA[i-1], n-1]$ and $T[\SA[i], n-1]$ (with $\LCP[0] = 0$). If we have the \FMindex{}, the LCP array, and the topology of the suffix tree, we get the \emph{compressed suffix tree}, which supports the full functionality of the suffix tree in a space-efficient manner \cite{Sadakane2007}.

\subsection{Graphs}\label{sect:graphs}

A \emph{graph} $G = (V, E)$ consists of a set of \emph{nodes} $V = \set{0, \dotsc, \abs{V}-1}$ and a set of \emph{edges} $E \subseteq V \times V$. We say that $(u, v) \in E$ is an edge \emph{from} node $u$ \emph{to} node $v$, and assume that the edges are \emph{directed}: $(u, v) \ne (v, u)$ for $u \ne v$. The \emph{indegree} $G.\gindegree(v)$ of node $v$ is the number of \emph{incoming} edges to $v$, while the \emph{outdegree} $G.\goutdegree(v)$ is the number of \emph{outgoing} edges from $v$.

The graphs we use are \emph{labeled} with alphabet $\Sigma$: each node $v \in V$ has a \emph{label} $G.\glabel(v) \in \Sigma$. A \emph{path} in a graph is a sequence of nodes $P = v_{0} \dotsm v_{\abs{P}-1}$ such that $(v_{i}, v_{i+1}) \in E$ for $0 \le i < \abs{P}-1$. We say that $v_{0}$ is the \emph{start} node and $v_{\abs{P}-1}$ is the \emph{end} node of the path. The label of path $P$ is the concatenation of node labels $G.\glabel(P) = G.\glabel(v_{0}) \dotsm G.\glabel(v_{\abs{P}-1})$.

If the graph has nodes with indegree or outdegree $0$, we add a \emph{source} node $s$ and a \emph{sink} node $t$ to it. To distinguish these technical nodes from the actual nodes, we label them with characters $G.\glabel(s) = \#$ and $G.\glabel(t) = \$$, which are not used anywhere else in the graph. We add an edge $(s, v)$ to all nodes $v \in V \setminus \set{s}$ with no incoming edges, and an edge $(v, t)$ from all nodes $v \in V \setminus \set{t}$ with no outgoing edges. We also add edge $(t, s)$ to guarantee that $G.\gindegree(v) \ge 1$ and $G.\goutdegree(v) \ge 1$ for all nodes $v \in V$. However, this edge is not considered a real edge, and no path can cross it.

We also consider infinite paths $P = (v_{i})_{i \in Z}$, generalizing the definition in a similar way as we did with infinite character sequences in Section~\ref{sect:strings}. We say that path $P$ is \emph{left-maximal} if it starts at the source node or extends infinitely to the left; \emph{right-maximal} if it ends at the sink node or extends infinitely to the right; and \emph{maximal} if it is both left-maximal and right-maximal.

Given a graph $G = (V, E)$, we may want to reason about the \emph{predecessors} of a node with the given label. Let $v \in V$ be a node and $c \in \Sigma$ be a character. We write $G.\gpred(v, c)$ to denote the set of nodes $u \in V$ such that $G.\glabel(u) = c$ and there is an edge $(u, v) \in E$.

We work with de~Bruijn graphs and their generalizations. For that purpose, we define collections of (finite or infinite) sequences suitable for constructing \orderk{k} de~Bruijn graphs.

\begin{Definition}[\kcollection{k}]
Let $\mathcal{S}$ be a collection of character sequences over alphabet $\Sigma$, and let $k > 0$ be a parameter value. We say that $\mathcal{S}$ is a \emph{\kcollection{k}}, if each sequence $S \in \mathcal{S}$
(a) is left-infinite or begins with $\#^{k}$;
(b) is right-infinite or ends with $\$^{k}$; and
(c) contains no other occurrences of characters $\#$ and $\$$.
\end{Definition}

\begin{Definition}[de~Bruijn graph]
Let $k > 0$, and let $\mathcal{S}$ be a \kcollection{k}. The \orderk{k} \emph{de~Bruijn graph} of $\mathcal{S}$ is a graph $G = (V, E)$ such that
\begin{itemize}
\item each node $v_{X} \in V$ represents a distinct \kmer{k} $X$ occurring in $\mathcal{S}$, with $G.\glabel(v_{X}) = X[0]$;
\item each node $v_{X} \in V$ has a \emph{key} $G.\gkey(v_{X}) = X$; and
\item each edge $(v_{X}, v_{Y}) \in E$ represents a \kmer{(k+1)} $X[0]Y = XY[k-1]$ occurring in $\mathcal{S}$.
\end{itemize}
We use $\#^{k}$ as the source node $s$ and $\$^{k}$ as the sink node $t$, adding the edge $(t, s)$ in the usual way.
\end{Definition}

De~Bruijn graphs have several properties that make them useful for indexing purposes. Node keys prefix-match the labels of all paths starting from the node. This makes it possible to sort the nodes unambiguously by path labels. Every substring of the \kcollection{k} is the label of a path in the de~Bruijn graph, and every path label of length at most $k+1$ is a substring of the collection. In Section~\ref{sect:path-indexes}, we develop an index structure based on a generalization of de~Bruijn graphs.

\subsection{FM-Index for Graphs}

Suffix trees, suffix arrays, and the \FMindex{} can be generalized to index multiple texts. There are also generalizations to other combinatorial structures. The \emph{XBW transform} \cite{Ferragina2009b} is an \FMindex{} for \emph{labeled trees}. The nodes of the tree are sorted by the path labels from the node to the root. $\BWT$ stores the labels of the children of each node, while leaf nodes are marked with special characters. If a node has $k$ children, we encode that as a binary sequence $0^{k-1} 1$. We concatenate these sequences to form the indegree bitvector $\bvIN$. The labels of the children of the $i$th node are found in $\BWT[\bvIN.\select(i, 1) + 1, \bvIN.\select(i + 1, 1)]$.

The \emph{generalized compressed suffix array} (GCSA) \cite{Siren2014} extends the XBW transform to a class of graphs that includes \emph{directed acyclic graphs} and de Bruijn graphs. Before indexing, we transform the graph into an equivalent graph, where the nodes can be sorted unambiguously by the labels of the right-maximal paths starting from them. The transformation increases the size of the graph exponentially in the worst case. In addition to sequences $\BWT$ and $\bvIN$, GCSA also uses an outdegree bitvector $\bvOUT$, which is encoded in the same way as $\bvIN$. \LFmapping{} uses $\select$ queries on bitvector $\bvIN$ to map nodes to BWT ranges, ordinary \LFmapping{} with $\BWT$ to map incoming edges to the corresponding outgoing edges, and $\rank$ queries on bitvector $\bvOUT$ to map the outgoing edges to the predecessor nodes.

\section{Path Indexes}\label{sect:path-indexes}

A \emph{path index} is a generalization of text indexes for \emph{labeled graphs}. Given a path index for \emph{input graph} $G = (V, E)$, we use the index to find the start nodes $v_{0} \in V$ of the paths $P = v_{0} \dotsm v_{\abs{X}-1}$ \emph{matching} pattern $X$ (paths $P$ with $G.\glabel(P) = X$).

The proofs of the lemmas can be found in Appendix~\ref{appendix:proofs}.

\subsection{De Bruijn Graphs as Path Indexes}

The \emph{\kmer{k} index} is the simplest path index. It consists of a set of \emph{key-value pairs} $(X, V_{X})$, where $X \in \Sigma^{k}$ is a \kmer{k} and $V_{X} \subseteq V$ is the set of the start nodes of the paths matching the \kmer{k}. If we store the pairs in a hash table, we can quickly search for patterns of length $k$. If we use binary search in a sorted list of pairs, queries become slower, but we gain the ability to search for patterns shorter than $k$ characters. The main drawback of these basic \kmer{k} indexes is their size, as they store the key-value pairs explicitly.

We can represent \kmer{k} indexes as de~Bruijn graphs. For that purpose, we define the de~Bruijn graph of graph $G = (V, E)$ by using the collection $\mathcal{S}$ of the labels of the maximal paths in the graph. If sequence $S \in \mathcal{S}$ is the label of path $P = (v_{i})_{i \in Z}$, we set $\mathcal{S}.\gnode(S, i) = v_{i}$ for all positions $i \in Z$. We transform $\mathcal{S}$ into a \kcollection{k} by inserting characters $\#$ to the beginning of each non-left-infinite sequence when necessary, and characters $\$$ to the end of each non-right-infinite sequence. If $S[i]$ is a $\#$ we inserted, we set $\mathcal{S}.\gnode(S, i) = s:j$, where $s \in V$ is the source node and $j$ is the distance to the nearest non-inserted $\#$ in $S$.

\begin{Definition}[de~Bruijn graph of a graph]\label{def:dbg} ~\\
Let $G$ be a labeled graph, and let $\mathcal{S}$ be the \kcollection{k} of maximal path labels in $G$. The \orderk{k} de~Bruijn graph of $\mathcal{S}$ is the \orderk{k} de~Bruijn graph of graph $G$.
\end{Definition}

\begin{figure*}[t!]
\includegraphics[width=\textwidth]{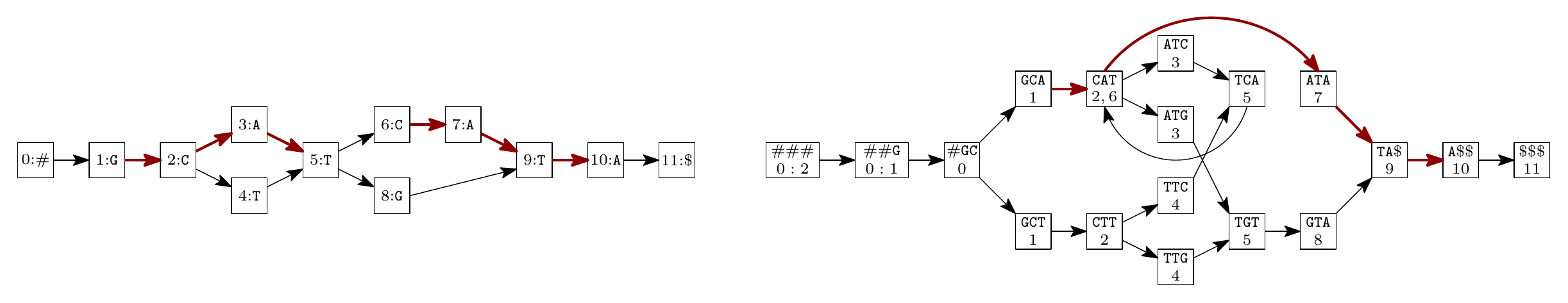}
\caption{Left: Input graph $G = (V, E)$, with each node $v \in V$ labeled with $v:G.\glabel(v)$. Right: The \protect\orderk{3} de~Bruijn graph $G' = (V', E')$ of graph $G$, with each node $v' \in V'$ labeled with $G'.\gkey(v')$ and $G'.\gvalue(v')$. Both: Edges $(t, s)$ are not shown. The highlighted path in the de~Bruijn graph is a false positive, as it consists of two disjoint paths in the input graph.}\label{figure:graph-dbg}
\end{figure*}

Let $G' = (V', E')$ be a de~Bruijn graph of graph $G = (V, E)$, and let $\mathcal{S}$ be the \kcollection{k} used to define it. We attach a set of nodes of graph $G$ to each node $v' \in V'$ as a \emph{value} $G'.\gvalue(v)$:
$$
\set{ \mathcal{S}.\gnode(S, i) \mid S \in \mathcal{S}, S[i, i+k-1] = G'.\gkey(v)}.
$$
Apart from some technicalities near the source/sink nodes, $G'.\gvalue(v)$ is the set of the start nodes of the paths matching pattern $G'.\gkey(v)$ in graph $G$. The index produces no \emph{false negatives} (path labels that exist in the input graph but not in the index). There may be \emph{false positives} (path labels that exist in the index but not in the input graph) with patterns longer than $k$, but we can avoid them by \emph{verifying} the results in the input graph. See Figure~\ref{figure:graph-dbg} for an example.

We can encode the de~Bruijn graph as a GCSA using $(\abs{V'} + \oh{\abs{V'}})(\log \sigma + 2)$ bits. By using a similar sampling scheme for the values as in the FM-index, we get a \kmer{k} index that uses a couple of bytes per \kmer{k} for typical variation graphs (see Section~\ref{sect:experiments}). While this is much less than with the basic \kmer{k} indexes, it is still too much for large variation graphs.

\subsection{Path Graphs}

When a path of length $k' < k$ has a unique label, its extensions become \emph{redundant} nodes in the \orderk{k} de~Bruijn graph, if the path branches after the first $k'$ characters. By using shorter keys when possible, we can build a smaller graph that is equivalent to the de~Bruijn graph as a path index.

\begin{Definition}[Path graph]
Let $k > 0$, let $\mathcal{S}$ be the \kcollection{k} of the labels of the maximal paths in graph $G = (V, E)$, and let $\mathcal{K}$ be a prefix-free set of substrings of length $k$ or less from $\mathcal{S}$. Assume that each right-maximal substring $S \in \mathcal{S}$ prefix-matches a string $K \in \mathcal{K}$ and that $\$^{k} \in \mathcal{S}$. The \orderk{k} \emph{path graph} of graph $G$ with \emph{key set} $\mathcal{K}$ is a graph $G' = (V', E')$, where
\begin{itemize}
\item each node $v_{K} \in V'$ represents a distinct key $K \in \mathcal{K}$, with $G'.\glabel(v_{K}) = K[0]$;
\item each node has a key $G'.\gkey(v_{K}) = K$ and a value $G'.\gvalue(v_{K}) = V_{K}$, where $V_{K}$ is the set of nodes $\mathcal{S}.\gnode(S, i) \in V$ for $S \in \mathcal{S}$ and positions $i$ such that $S[i, i+\abs{K}-1] = K$; and
\item each edge $(v_{K}, v_{K'}) \in E'$ represents the occurrence of substring $K[0] K'$ in $\mathcal{S}$ such that strings $K$ and $K[0] K'$ prefix-match.
\end{itemize}
We use $\#^{k}$ as the source node $s$ and $\$^{k}$ as the sink node $t$, adding the edge $(t, s)$ in the usual way.
\end{Definition}

\begin{Definition}[Path graph as an index] ~\\
Let $G = (V, E)$ be a graph, and let $G' = (V', E')$ be a path graph of $G$.
\begin{itemize}
\item Pattern $X \in \Sigma^{\ast}$ \emph{matches} node $v \in V'$, if there is a path $P'$ in $G'$ with $G'.\glabel(P') = X$. We use $G'.\find(X)$ to denote the set of nodes $V'_{X} \subseteq V'$ matching the pattern.
\item If $V'_{X} \subseteq V'$ is the set of nodes matching pattern $X$, the set of \emph{occurrences} for the pattern is $G'.\locate(V'_{X}) = \bigcup_{v' \in V'_{X}} G'.\gvalue(v')$. We use $G'.\locate(X)$ to denote $G'.\locate(G'.\find(X))$.
\end{itemize}
\end{Definition}

\begin{lemma}[No false negatives]\label{lemma:pg-fn}
Let $G' = (V', E')$ be a path graph of $G = (V, E)$, and let $X \in \patternset$ be a pattern. Set $G'.\locate(X)$ contains the start nodes of all paths $P$ in graph $G$ with $G.\glabel(P) = X$.
\end{lemma}

\begin{lemma}[Context length]\label{lemma:pg-context}
Let $G' = (V', E')$ be a path graph, and let $X \in \patternset$ be a pattern. Set $G'.find(X)$ consists of all nodes $v' \in V'$ such that $X[0, m-1]$ is a prefix of $G'.\gkey(v')$, for a context length $m$, which depends on the graph and the pattern.
\end{lemma}

\begin{lemma}[Short keys]\label{lemma:pg-keys}
Let $G' = (V', E')$ be a path graph with $\abs{G'.\gkey(u')} \le \abs{G'.\gkey(v')}+1$ for all edges $(u', v') \in E'$. Then
(a) $\abs{G'.\gpred(v', c)} \le 1$ for all nodes $v' \in V'$ and characters $c \in \Sigma$; and
(b) key $G'.\gkey(v')$ prefix-matches pattern $X \in \patternset$ for all nodes $v' \in G'.\find(X)$.
\end{lemma}

A path graph may produce false positives with patterns longer than $k'$ characters, where $k'$ is the length of the shortest key. In the next section, we define a class of path graphs that can be proven to be equivalent to de~Bruijn graphs.

\subsection{Pruned de Bruijn Graphs}

We can compress de~Bruijn graphs structurally by merging keys sharing a common prefix, if the corresponding values are identical. These pruned de~Bruijn graphs, which arise naturally from GCSA construction, are similar to manifold de Bruijn graphs \cite{Lin2014}. As path indexes, they are equivalent to de~Bruijn graphs with patterns of length up to $k$ characters.

\begin{Definition}[Equivalent path graphs]
Let $G'$ and $G''$ be two path graphs, and let $k > 0$ be a parameter value. We say that graphs $G'$ and $G''$ are \emph{\kequivalent{k}}, if we have $G'.\locate(X) = G''.\locate(X)$ for all patterns $X \in \patternset$ with $1 \le \abs{X} \le k$.
\end{Definition}

\begin{Definition}[Pruned de~Bruijn graph]
Let $G$ be a graph, and let $G'$ be an \orderk{k} path graph of $G$. Path graph $G'$ is an \orderk{k} \emph{pruned de~Bruijn graph}, if it is \kequivalent{k} to the \orderk{k} de~Bruijn graph of $G$.
\end{Definition}

\begin{lemma}[No short false positives]\label{lemma:dbg-fp} ~\\
Let $G = (V, E)$ be a graph, let $G' = (V', E')$ be an \orderk{k} pruned de~Bruijn graph of graph $G$, and let $X \in \patternset$ be a pattern with $1 \le \abs{X} \le k$. Then $G'.\locate(X)$ is a set of start nodes $v\in V$ of paths matching the pattern in graph $G$.
\end{lemma}

\begin{lemma}[Pruning]\label{lemma:dbg-prune}
Let $G = (V, E)$ be a graph, let $G' = (V', E')$ be the \orderk{k} pruned de~Bruijn graph of $G$ with key set $\mathcal{K}$, let $K \in \Sigma^{\ast}$ be a string of length $\abs{K} > 0$, and let $V'_{K}$ be the set of nodes $v' \in V'$ having string $K$ as a proper prefix of $G'.\gkey(v')$.
If $\abs{V'_{K}} > 0$ and $G'.\gvalue(u') = G'.\gvalue(v')$ for all $u', v' \in V'_{K}$, the path graph with key set $(\mathcal{K} \setminus \set{ G'.\gkey(v') \mid v' \in V'_{K}}) \cup \set{K}$ is an \orderk{k} pruned de~Bruijn graph of $G$.
\end{lemma}

\begin{figure*}[t!]
\includegraphics[width=\textwidth]{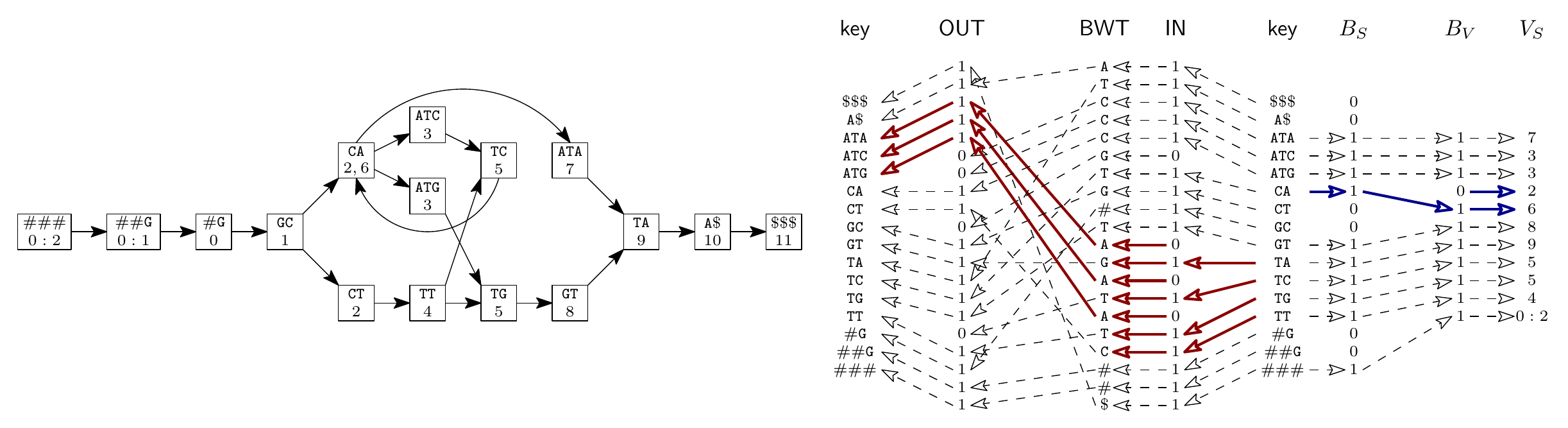}
\caption{Left: An \protect\orderk{3} pruned de~Bruijn graph $G''$ \protect\kequivalent{3} to the de~Bruijn graph in Figure~\ref{figure:graph-dbg}. Right: GCSA for graph $G''$. Leftward arrows illustrate backward searching, with the red arrows showing it from $\mathtt{T}$ to $\mathtt{AT}$. Rightward arrows mark the samples belonging to each node, with the blue ones showing them for node $\mathtt{CAT}$.}\label{figure:pruned-index}
\end{figure*}

We can compress a de~Bruijn graph structurally by merging sets of nodes sharing a common prefix of their keys, as long as the conditions of Lemma~\ref{lemma:dbg-prune} hold. Let $G' = (V', E')$ be an \orderk{k} pruned de~Bruijn graph, and let $G'' = (V'', E'')$ be the same graph after further pruning. Each node $v'' \in V''$ is an \emph{equivalence class} of nodes $V'(v'') \subseteq V'$ corresponding to a shared prefix $G''.\gkey(v'')$ of keys. For all $v' \in V'(v'')$, we have $G'.\gvalue(v') = G''.\gvalue(v'')$. See Figure~\ref{figure:pruned-index} for an example of a pruned de~Bruijn graph.

\begin{Definition}[Maximally pruned graph] ~\\
Let $G'$ be a a pruned de~Bruijn graph of graph $G$. We say that $G'$ is \emph{maximally pruned}, if we cannot prune it any further using Lemma~\ref{lemma:dbg-prune}.
\end{Definition}

\begin{lemma}[Maximal pruning]\label{lemma:dbg-maximal}
Let $G' = (V', E')$ be a maximally pruned de~Bruijn graph of $G = (V, E)$. Then $\abs{G'.\gkey(u')} \le \abs{G'.\gkey(v')}+1$ for all $(u', v') \in E'$.
\end{lemma}

\section{GCSA2}

As in the original GCSA, we sort the nodes of the path graph in lexicographic order, encode the indegrees and outdegrees in bitvectors $\bvIN$ and $\bvOUT$, and store the predecessor labels in $\BWT$. See Figure~\ref{figure:pruned-index} for an example. If lexicographic range $[sp_{i+1}, ep_{i+1}]$ matches suffix $X[i+1, \abs{X}-1]$ of pattern $X$, we can find the range $[sp_{i}, ep_{i}]$ matching suffix $X[i, \abs{X}-1]$ as
\begin{align*}
[sp_{in}, ep_{in}] = [&\bvIN.\select(sp_{i+1}, 1) + 1, \\
 & \bvIN.\select(ep_{i+1}+1, 1)]; \\
[sp_{out}, ep_{out}] = [&\LF(sp_{in}, X[i]), \\
 & \LF(ep_{in}+1, X[i]) - 1]; \\
[sp_{i}, ep_{i}] = [&\bvOUT.\rank(sp_{out}, 1), \\
 & \bvOUT.\rank(ep_{out}, 1)].
\end{align*}

In order to support $\locate$ queries, we \emph{sample} the values of a node, if the node has multiple incoming edges or the values cannot be derived from the predecessor. We may also sample other values to improve query performance. The sampled nodes are marked in bitvector $B_{S}$, the number of values in each sample is encoded in unary in bitvector $B_{V}$, and the sampled values are stored in integer array $V_{S}$. A detailed description of the data structure can be found in Appendix~\ref{appendix:gcsa2}.

GCSA2 construction starts from paths of length $k$ in the input graph. We build a maximally pruned \orderk{2k}, \orderk{4k}, or \orderk{8k} de~Bruijn graph using a similar \emph{prefix-doubling} algorithm as in the original GCSA \cite{Siren2014}, and encode the result as a GCSA. To avoid excessive memory usage, we keep the paths and the graphs on disk, and read only a single chromosome at a time into memory. The details of the construction algorithm can be found in Appendix~\ref{appendix:construction}.

We can improve the query performance with (maximally pruned) de~Bruijn graphs by using a \emph{simplified encoding} (Appendix~\ref{appendix:encoding}). We replace $\BWT$ and $\bvIN$ with bitvectors $B_{c}$ for all $c \in \Sigma$, where $B_{c}[j] = 1$ if and only if node $j$ in lexicographic order has a predecessor with label $c$. This simplifies backward searching to
\begin{align*}
sp_{out} & = C[X[i]] + B_{X[i]}.\rank(sp_{i+1}, 1); \\
ep_{out} & = C[X[i]] + B_{X[i]}.\rank(ep_{i+1}+1, 1) - 1; \\
[sp_{i}, ep_{i}] & = [\bvOUT.\rank(sp_{out}, 1), \bvOUT.\rank(ep_{out}, 1)].
\end{align*}

The compacted trie of keys resembles a suffix tree. We can simulate it space-efficiently by using the LCP array \cite{Abouelhoda2004,Fischer2009a}, and thus extend GCSA2 to support many suffix tree operations. For example, we can search for \emph{maximal exact matches} by using \LFmapping{} and $\parent$ queries \cite{Ohlebusch2010a}, and use that as a basis for a read aligner similar to BWA-MEM \cite{Li2013}. We can also use \emph{document counting} techniques \cite{Sadakane2007a} to quickly count the number of distinct matches in a lexicographic range. Further details of these extensions can be found in Appendix~\ref{appendix:extensions}.

\section{Implementation and Experiments}\label{sect:experiments}

GCSA2 is the path indexing library of vg \cite{Garrison2014-2016}. The implementation is written in C++, and the source code is available on GitHub.\footnote{\url{https://github.com/jltsiren/gcsa2}} It depends on \emph{SDSL} \cite{Gog2014b} and \emph{libstdc++ parallel mode}. We use the simplified encoding (Appendix~\ref{appendix:encoding}) with fast non-compressed bitvectors in most index components. Bitvectors $B_{c}$ for rare characters ($\baseN$, $\#$, and $\$$) are compressed as sparse bitvectors \cite{Okanohara2007}.

We used a system with two 16\nobreakdash-core AMD Opteron 6378 processors and 256~gigabytes of memory for the experiments, and stored all files on a distributed Lustre file system. The system was running Ubuntu~12.04 on Linux kernel~3.2.0. We used vg version~1.3.0 for processing the graphs and GCSA2 version~0.8 using SDSL version~2.1.1 for the benchmarks. All code was compiled with gcc/g++ version~4.9.2.

\subsection{Construction}

\emph{Variation graphs}, as defined in vg, use strings as node labels. A node can be traversed in both forward and reverse complement orientations, and edges may cross between the orientations. For indexing, the graph is implicitly converted into an input graph with single-character labels. We always sample the input graph nodes corresponding to the initial offsets of variation graph node labels.

We built vg graphs from the human reference genome (GRCh37) and 1000~Genomes Project variation \cite{1000GP2015}. To avoid excessive growth, we removed paths where \kmer{16}s crossed more than $4$ nontrivial edges with \texttt{vg mod -p -l -e 4}, and subgraphs shorter than $100$ bases with \texttt{vg mod -S -l 100}. We extracted all paths of length $16$ from the forward strand of the graph. There were a total of 4.80~billion paths with 1.53~billion distinct labels. We then built GCSA with 1\nobreakdash--3 doubling steps, producing \orderk{32}, \orderk{64}, and \orderk{128} indexes.

Tables~\ref{table:construction} and \ref{table:indexes} show construction requirements and index sizes, respectively. We can build a whole-genome index overnight using less than 96~gigabytes of memory, including disk cache. The index contains $1.031^{k} \cdot 2.348$~billion \kmer{k}s, but the path graph only uses 4.4\nobreakdash--5.7~billion nodes to represent them. For $k = 128$, GCSA2 requires 0.63~bits per \kmer{k}, out of which 0.28~bits is used for the path graph. Extensions based on suffix trees increase the size to 1.08~bits per \kmer{k}.

\begin{table}
\begin{center}
\begin{tabular}{c|ccc|cc}
\hline
$k$ & \textbf{Time} & \textbf{Memory} & \textbf{Disk} & \textbf{Read} & \textbf{Write} \\
\hline
 32 & 7.44 h & 59.8 GB & 387 GB & 1.37 TB & 0.88 TB \\
 64 & 10.4 h & 51.9 GB & 415 GB & 2.03 TB & 1.51 TB \\
128 & 14.1 h & 52.3 GB & 478 GB & 2.78 TB & 2.25 TB \\
\hline
\end{tabular}
\caption{GCSA2 construction. Order of the path graph; construction time in hours; peak memory and disk usage in gigabytes; and disk I/O volume for reading and writing in terabytes.}\label{table:construction}
\end{center}
\end{table}

\begin{table*}[t]
\begin{center}
\begin{tabular}{c|cc|c|c|c}
\hline
$k$ & \textbf{\kmer{k}s} & \textbf{Nodes} & \textbf{Graph} & \textbf{Index} & \textbf{With extensions} \\
\hline
 32 & 6.20G & 4.37G & 2.89 GB / 4.00 bits & 9.50 GB / 13.2 bits & 13.2 GB / 18.2 bits \\
 64 & 16.7G & 5.24G & 3.46 GB / 1.78 bits & 8.64 GB / 4.46 bits & 13.6 GB / 6.99 bits \\
128 &  116G & 5.73G & 3.78 GB / 0.28 bits & 8.58 GB / 0.63 bits & 14.6 GB / 1.08 bits \\
\hline
\end{tabular}
\caption{GCSA2 index sizes. Order of the path graph; number of \kmer{k}s and nodes in the path graph in billions; index size in gigabytes and in bits per \kmer{k} for the graph ($B_{c}$ and $\bvOUT$), the index (the graph, $B_{S}$, $B_{V}$, and $V_{S}$), and the index with the extensions from Appendix~\ref{appendix:extensions}.}\label{table:indexes}
\end{center}
\end{table*}

Index construction uses more memory with $k = 32$ than with larger values of $k$. The \orderk{32} path graph has more nodes, where we cannot derive the values from the predecessor node. As we sample more values, we need more memory in the final phase of construction. With larger values of $k$, the path graph resembles the input graph better, and we sample less values. For the same reason, index size decreases with larger values of $k$, even though the graph requires more space.

\subsection{Queries}

We compared the query performance of the \orderk{128} GCSA2 to several FM-indexes for the reference sequence. SSA is the SDSL implementation (\texttt{csa\_wt<>}) of the \emph{succinct suffix array} \cite{Maekinen2005}, using a Huffman-shaped wavelet tree on top of the BWT. As the default FM-index in SDSL, it prioritizes query performance over compression. We used SSA with SA sample period 17 for good $\locate$ performance. BWA is the FM-index in the \emph{Burrows-Wheeler Aligner} \cite{Li2009} (version~0.7.15 with the default SA sample period 32). Optimized for DNA sequences, BWA indexes both the reference and its reverse complement.

As building the original GCSA requires around $65n$ bytes of memory for a path graph with $n$ nodes, we could not compare GCSA and GCSA2 directly on a system with 256~gigabytes of memory. Instead, we used \emph{RLCSA} \cite{Maekinen2010} (May~2016 version) as a proxy. The RLCSA is an FM-index for repetitive sequence collections using the same basic components as the original GCSA. Under a mixed query load, RLCSA with SA sample period 32 is 1.5x to 3x faster than GCSA, depending on algorithmic overhead and the mix of $\find$ and $\locate$ queries \cite{Siren2014}.

We extracted \kmer{k}s for $k \in \set{16, 32, 64, 128}$ from the (non-pruned) vg graphs by using \texttt{vg sim}, filtered out \kmer{k}s consisting entirely of $\baseN$s, and queried for the remaining \kmer{k}s using a single thread. The results can be seen in Table~\ref{table:benchmark}.

\begin{table*}[t]
\begin{center}
\begin{tabular}{cc|cccc|cccc}
\hline
$k$ & \textbf{Patterns} & \textbf{Index} & \textbf{Found} & \textbf{Nodes} & \textbf{Occs} & $\find()$ & $\parent()$ & $\countq()$ & $\locate()$ \\
\hline
 16 & 351584 & GCSA2 & 347453 & 2477M &  872M & 4.75 \textmu{}s & 0.42 \textmu{}s & 0.87 \textmu{}s & 5.85 \textmu{}s \\
    &        & SSA   & 301538 &    -- &  782M & 6.00 \textmu{}s &              -- &              -- & 2.43 \textmu{}s \\
    &        & BWA   & 320764 &    -- & 1564M & 3.64 \textmu{}s &              -- &              -- & 4.65 \textmu{}s \\
    &        & RLCSA & 301538 &    -- &  782M & 23.7 \textmu{}s &              -- &              -- & 8.12 \textmu{}s \\
\hline
 32 & 351555 & GCSA2 & 333258 &  112M & 34.3M & 10.8 \textmu{}s & 0.28 \textmu{}s & 0.38 \textmu{}s & 5.44 \textmu{}s \\
    &        & SSA   & 153957 &    -- & 26.6M & 10.9 \textmu{}s &              -- &              -- & 2.16 \textmu{}s \\
    &        & BWA   & 156080 &    -- & 52.9M & 6.57 \textmu{}s &              -- &              -- & 3.19 \textmu{}s \\
    &        & RLCSA & 153957 &    -- & 26.6M & 47.6 \textmu{}s &              -- &              -- & 5.87 \textmu{}s \\
\hline
  64 & 351567 & GCSA2 & 326101 & 2.63M & 1.35M & 22.5 \textmu{}s & 0.26 \textmu{}s & 0.29 \textmu{}s & 2.92 \textmu{}s \\
     &        & SSA   &  88184 &    -- & 0.84M & 17.1 \textmu{}s &              -- &              -- & 1.89 \textmu{}s \\
     &        & BWA   &  88786 &    -- & 1.60M & 10.3 \textmu{}s &              -- &              -- & 2.34 \textmu{}s \\
     &        & RLCSA &  88184 &    -- & 0.84M & 74.3 \textmu{}s &              -- &              -- & 5.97 \textmu{}s \\
\hline
 128 & 351596 & GCSA2 & 316500 & 0.32M & 0.37M & 45.3 \textmu{}s & 0.26 \textmu{}s & 0.26 \textmu{}s & 3.13 \textmu{}s \\
     &        & SSA   &  35678 &    -- & 0.08M & 23.5 \textmu{}s &              -- &              -- & 3.47 \textmu{}s \\
     &        & BWA   &  35741 &    -- & 0.12M & 14.0 \textmu{}s &              -- &              -- & 3.46 \textmu{}s \\
     &        & RLCSA &  35678 &    -- & 0.08M & 91.7 \textmu{}s &              -- &              -- & 12.9 \textmu{}s \\
\hline
\end{tabular}
\caption{Query benchmarks using an \orderk{128} GCSA2 and various FM\nobreakdash-indexes. Pattern length; number of patterns; index type; number matching patterns, matching nodes, and distinct occurrences; average time for $[sp, ep] = \find(X)$, $\parent(sp, ep)$, and $\countq(sp, ep)$ queries in microseconds; and average time per value for $\locate(sp, ep)$ queries in microseconds.}\label{table:benchmark}
\end{center}
\end{table*}

Backward searching in an FM-index stops early if there are no matches. In order to compare the $\find$ performance of the indexes reliably, we must hence concentrate on the \kmer{16}s, where the fraction of matching patterns is similar for all indexes. GCSA2 and the fast FM-indexes (SSA and BWA) all have similar performance, while RLCSA is several times slower. As a result, we can estimate that $\find$ queries in GCSA2 are an order of magnitude faster than in GCSA.

When comparing the $\locate$ performance of different FM-indexes, the distribution of the query positions should be close to uniform. Otherwise the biases from e.g.~different suffix array sampling strategies or the variation in the number of distinct occurrences per node in GCSA2 can make the results unreliable. As the \kmer{k}s have been sampled uniformly from the variation graph, we get the best results with the \kmer{16}s, where all indexes can match most of the patterns.

GCSA2 uses denser SA sampling than the other indexes, with effective sample period 10.6. On the average, GCSA2 calls $\locate$ for 2.84~nodes per distinct value, making the amount of work comparable to sample period 30.2. SSA with sample period 17 is 2.4x faster than GCSA2, mostly because it has to do less work. BWA with sample period 32 is closer to GCSA2 in $\locate$ performance. RLCSA is slower than the other indexes, but the difference is smaller than with $\find$ queries due to the optimizations for retrieving suffix array ranges. Assuming that $\locate$ queries are 3x slower in GCSA than in RLCSA, as GCSA does not use the optimizations, we can estimate that GCSA2 is 4x faster than GCSA.

The remaining queries, $\parent$ and $\countq$, take a fraction of a microsecond. As a $\parent$ query takes comparable time to a single step of backward searching, it will not be a bottleneck in finding maximal exact matches. Counting the number of distinct occurrences with a $\countq$ query is faster than retrieving even a single occurrence.

\section{Discussion}

GCSA2 is a path index for variation graphs. It uses a de~Bruijn graph as a \kmer{k} index of the variation graph, prunes it by merging redundant subgraphs, and encodes the result with a generalization of the FM\nobreakdash-index. The index supports queries of length up to $k$ exactly, and longer queries with false positives. GCSA2 also includes extensions based on suffix trees; other extensions have been considered but not implemented (see Appendix~\ref{appendix:hypertext}). The index is used in the variation graph toolkit vg for e.g.~read alignment based on maximal exact matches.

We can build a whole-genome index overnight on a system with 96~gigabytes of memory and a few hundred gigabytes of fast disk space. The resulting index takes less than 15~gigabytes, or 1.08~bits per \kmer{k} for the \orderk{128} index with extensions. Query performance is comparable to that of fast FM-indexes for sequences.

The primary design goals for GCSA2 were query performance and index size. The index works with arbitrary graphs, supports queries that are long enough to map short reads in one piece without false positives, and provides several options for dealing with complex regions. Other path indexes work with a more restricted class of graphs \cite{Siren2014,Huang2013,Kim2015-2016,Maciuca2016}, are at least an order of magnitude slower \cite{Huang2013,Maciuca2016}, require much more space \cite{Bowe2012,Pell2012,Cazaux2014,Marcus2014}, or are theoretical proposals that have never been implemented \cite{Thachuk2013}.

We may want to determine whether a pattern matches known \emph{haplotypes} or only their recombinations. As GCSA2 does not support this directly, vg must determine it afterwards using a separate structure \cite{Novak2016}. The \emph{FM\nobreakdash-index of alignment} \cite{Na2015,Na2016} embeds the haplotypes directly in a GCSA-like index and reports the haplotypes matching the $\find$ query. While the solution depends on specific properties of the graph, it could be possible to extend it to work with any GCSA.


\paragraph{Acknowledgements.} The author thanks Erik Garrison, Richard Durbin, and Adam Novak for the fruitful discussions while developing the GCSA2 index.

\appendix
\section{Proofs of Lemmas}\label{appendix:proofs}

\begin{proof}[Lemma~\ref{lemma:pg-fn}: No false negatives]
Let $\mathcal{S}$ be the \kcollection{k} used for building the path graph, and let $P$ be a path starting from $v_{0} \in V$ with $G.\glabel(P) = X$. The collection contains a sequence $S \in \mathcal{S}$ such that $S[i, i+\abs{X}-1] = X$ and $\mathcal{S}.\gnode(S, i) = v_{0}$.

For all positions $j$ with $i \le j \le i+\abs{X}-1$, there is a node $v'_{j} \in V'$ with $G'.\glabel(v'_{j}) = S[j]$ and $G'.\gkey(v'_{j}) = S[j, j+\abs{G'.\gkey(v'_{j})}-1]$.
By definition, path graph $G'$ has an edge $(v'_{j}, v'_{j+1}) \in E'$ for all such positions $j$.
Hence $P' = v'_{i} \dotsm v'_{i+\abs{X}-1}$ is a path in $G'$ with $G'.\glabel(P') = X$.
As path $P'$ starts from node $v'_{i} \in V'$, node $v'_{i}$ is included in the set $G'.\find(X)$.
Furthermore, $v_{0} = \mathcal{S}.\gnode(S, i) \in G'.\gvalue(v'_{i}) \subseteq G'.\locate(X)$.
\end{proof}

\begin{proof}[Lemma~\ref{lemma:pg-context}: Context length]
If $\abs{X} \le 1$, the statement is true by definition for $m = \abs{X}$. Now let $M_{i+1} = G'.\find(X[i+1, \abs{X}-1])$ be the set of all nodes $v' \in V'$ such that substring $X[i+1, i+m_{i+1}]$ is a prefix of key $G'.\gkey(v')$, and assume that the set is nonempty.

Consider the set
$$
M_{i} = \bigcup_{v' \in M_{i+1}} G'.\gpred(v', X[i]) = G'.\find(X[i, \abs{X}-1]).
$$
There is an edge $(u', v') \in E'$ if and only if key $G'.\gkey(u')$ prefix-matches string $G'.\glabel(u') \cdot G'.\gkey(v')$. Hence key $G'.\gkey(u')$ prefix-matches string $X[i, i+m_{i+1}]$ for all nodes $u' \in M_{i}$.

Now let $u' \in V'$ be a node with key $G'.\gkey(u')$ prefix-matching string $X[i, i+m_{i+1}]$. If the key is a prefix of string $X[i, i+m_{i+1}]$, there is an edge $(u', v') \in E'$ to all nodes $v' \in M_{i+1}$, and hence $u' \in M_{i}$. Otherwise let $S[j, j+\abs{G'.\gkey(u')}-1] = G'.\gkey(u')$ be a substring of the \kcollection{k} used for building the path graph. As $S[j+1, j+m_{i+1}] = X[i+1, i+m_{i+1}]$, the substring starting at $S[j+1]$ is represented by a node $v' \in M_{i+1}$, and hence $u' \in M_{i}$.

Set $M_{i}$ is the set of all nodes $u' \in V'$ such that substring $X[i, i+m_{i+1}]$ prefix-matches key $G'.\gkey(u')$. If $\abs{M_{i}} > 1$, string $X[i, i+m_{i+1}]$ is a proper prefix of key $G'.\gkey(u')$ for all nodes $u' \in M_{i}$ due to the prefix-free property, and we can set $m_{i} = m_{i+1}+1$. Otherwise we set $m_{i} = \min(m_{i+1}+1, \abs{G'.\gkey(u')})$ for the only node $u' \in M_{i}$.
\end{proof}

\begin{proof}[Lemma~\ref{lemma:pg-keys}: Short keys]
(a) If node $v'$ has multiple predecessors with label $c$, the keys of the predecessors must be longer than string $c \cdot G'.\gkey(v')$, as the key set is prefix-free.

(b) By the construction in the proof of Lemma~\ref{lemma:pg-context}, the context length for pattern $X$ in graph $G'$ is $\min(\abs{X}, \abs{G'.\gkey(v')})$.
\end{proof}

\begin{proof}[Lemma~\ref{lemma:dbg-fp}: No short false positives]
We may assume without loss of generality that graph $G'$ is a de~Bruijn graph. Let $v' \in G'.\find(X)$ be a node. By Lemma~\ref{lemma:pg-keys}, pattern $X$ prefix-matches key $G'.\gkey(v')$. For every node $v \in G'.\gvalue(v')$, there is a substring $S[i, i+k-1] = G'.\gkey(v'_{0})$ in the \kcollection{k} $\mathcal{S}$ used for building graph $G'$, with $\mathcal{S}.\gnode(S, i) = v$. Prefix $S[i, i+\abs{X}-1] = X$ of the substring corresponds to a path with label $X$ starting from node $v$ in graph $G$.
\end{proof}

\begin{Definition}[Equivalent paths] ~\\
Let $G'$ and $G''$ be path graphs of the same graph, and let $P' = v'_{0} \dotsm v'_{n-1}$ and $P'' = v'_{0} \dotsm v''_{n-1}$ be paths in graphs $G'$ and $G''$, respectively. We say that paths $P'$ and $P''$ are \emph{equivalent}, if for $0 \le i < n$, keys $G'.\gkey(v'_{i})$ and $G''.\gkey(v''_{i})$ have a common prefix $K_{i}$ such that $G'.\gvalue(v') = G''.\gvalue(v'')$ for all nodes $v' \in V'$ and $v'' \in V''$ having $K_{i}$ as a prefix of their keys.
\end{Definition}

\begin{proof}[Lemma~\ref{lemma:dbg-prune}: Pruning]
Let $G'' = (V'', E'')$ be the path graph corresponding to the new key set, let $v''_{K} \in V''$ be the node with key $K$, and let $\mathcal{S}$ be the \kcollection{k} used to define the path graphs.

Consider the edge $(u', v') \in E'$ defined by substring $S'$ of $\mathcal{S}$. The same substring also defines an edge $(u'', v'') \in E''$, where either (a) $G''.\gkey(u'') = G'.\gkey(u')$ or (b) $u'' = v''_{K}$ and $u' \in V'_{K}$, and the same holds for nodes $v''$ and $v'$. We can hence transform any path in graph $G'$ into an equivalent path in graph $G''$ by replacing nodes $v' \in V'_{K}$ with node $v''_{K}$.

Let $P'' = v''_{0} \dotsm v''_{\abs{P''}-1}$ be a in graph $G''$. We transform it into an equivalent path $P' = v'_{0} \dotsm v'_{\abs{P''}-1}$ in graph $G'$. There are two cases for $\abs{P''} = 1$. If $v''_{0} = v''_{K}$, we can replace it with any $v'_{0} \in V'_{K}$. Otherwise we use the node $v'_{0} \in V'$ with $G'.\gkey(v'_{0}) = G''.\gkey(v''_{0})$.

In the general case $\abs{P''} > 1$, assume that we have transformed the suffix $v''_{1} \dotsm v''_{\abs{P''}-1}$ of path $P''$ into an equivalent path $v'_{1} \dotsm v'_{\abs{P''}-1}$ in graph $G'$. Because $G'.\gvalue(v'_{1}) = G''.\gvalue(v''_{1})$, node $v'_{1}$ must have a predecessor with label $c = G''.\glabel(v''_{0})$. We can choose any such predecessor as node $v'_{0}$.

Consider the predecessors $v' \in G'.\gpred(v'_{1}, c)$ and $v'' \in G''.\gpred(v''_{1}, c)$. Their keys prefix-match string $Y = c \cdot K_{1}$. There are three cases:
\begin{enumerate}
\item If $Y$ is a prefix of $G'.\gkey(v')$, the key $G'.\gkey(w')$ of every successor $w'$ of node $v'$ prefix-matches $K_{1}$ and hence has $K_{1}$ as a prefix. Therefore $G'.\gvalue(v')$ is the union of sets $G.\gpred(v, c)$ over $v \in G'.\gvalue(v'_{1})$. If $Y$ is also a prefix of $G''.\gkey(v'')$, nodes $v'$ and $v''$ have identical value sets by the same reasoning.

\item If $Y$ is a prefix of $G'.\gkey(v')$ and $G''.\gkey(v'')$ is a proper prefix of $Y$, there is only one possible predecessor $v'' = v''_{0}$. Hence $v' \in V'_{K}$ and $v''_{0} = v''_{K}$.

\item If key $G'.\gkey(v')$ is a proper prefix of string $Y$, there is only one possible predecessor $v' = v'_{0}$. Because key $G''.\gkey(v''_{0})$ prefix-matches string $Y$, it must also be a proper prefix of the string. Hence either $G'.\gkey(v'_{0}) = G''.\gkey(v''_{0})$ or $v'_{0} \in V'_{K}$ and $v''_{0} = v''_{K}$.
\end{enumerate}
In every case, $G'.\gvalue(v'_{0}) = G''.\gvalue(v''_{0})$, and we can use shorter of strings $Y$ and $G''.\gkey(v''_{0})$ as $K_{0}$.

We can transform any path $P'$ in graph $G'$ into an equivalent path $P''$ in graph $G''$, and the other way around. Because the labels of equivalent paths and the value sets of their start nodes are identical, we have $G'.\locate(X) = G''.\locate(X)$ for all patterns $X \in \patternset$ with $\abs{X} > 0$.
\end{proof}

\begin{proof}[Lemma~\ref{lemma:dbg-maximal}: Maximal pruning]
Let $G_{d} = (V_{d}, E_{d})$ the de~Bruijn graph of graph $G$ with the same order $k$ as graph $G'$, and let $\mathcal{S}$ be the \kcollection{k} used for building the path graphs. If $v' \in V'$ is a node, then $G_{d}.\gvalue(v) = G'.\gvalue(v')$ for all nodes $v \in V_{d}(v')$.

Assume that $\abs{G'.\gkey(u')} > \abs{G'.\gkey(v')}+1$ for an edge $(u', v') \in E'$. String $G'.\glabel(u') \cdot G'.\gkey(v')$ must then be a prefix of key $G'.\gkey(u')$. There cannot be edges $(u', w')$ to other nodes $w' \ne v'$, as keys $G'.\gkey(v')$ and $G'.\gkey(w')$ would prefix-match.

Let $S[i, i+\abs{G'.\gkey(u')}-1] = G'.\gkey(u')$ be a substring of $\mathcal{S}$. Because $G'.\gkey(v')$ is a substring of $G'.\gkey(u')$ and $G'$ is a pruned de~Bruijn graph, the set of nodes $\mathcal{S}.\gnode(S, i+1)$ over all occurrences of substring $G'.\gkey(u')$ in $\mathcal{S}$ is $G'.\gvalue(v')$. As node $u'$ has no other successors, set $G'.\gvalue(u')$ is the union of sets $G.\gpred(v, G'.\glabel(u'))$ over all nodes $v \in G'.\gvalue(v')$.

The above is true for all $x' \in G'.\gpred(v', G'.\glabel(u'))$. Hence we can prune graph $G'$ further using string $G'.\glabel(u') \cdot G'.\gkey(v')$ as the new key in Lemma~\ref{lemma:dbg-prune}.
\end{proof}

\section{GCSA for Path Graphs}\label{appendix:gcsa2}

Let $G' = (V', E')$ be a path graph. We sort the nodes $V'$ by their keys in lexicographic order and generate the sequences $\BWT$, $\bvIN$, and $\bvOUT$ from the nodes in that order. For each node $v' \in V'$, we append $\BWT$ with the predecessor labels $G.\glabel(u')$ for all edges $(u', v') \in E'$; $\bvIN$ with the indegree encoded as $0^{G'.\gindegree(v')-1} 1$; and $\bvOUT$ with the outdegree as $0^{G'.\goutdegree(v')-1} 1$.

If node $v' \in V'$ has lexicographic rank $i$, the range of incoming edges $(u', v') \in E'$ to that node is $[sp_{in}, ep_{in}] = [\bvIN.\select(i, 1) + 1, \bvIN.\select(i+1, 1)]$. The labels of the predecessor nodes are encoded in $\BWT[sp_{in}, ep_{in}]$. Sorting the incoming edges by pairs $(\BWT[j], i)$, where $\BWT[j]$ corresponds to edge $(u', v') \in E'$, is equivalent to sorting them by strings $G'.\glabel(u') \cdot G'.\gkey(v')$. As multiple edges may have the same sort key, our sorting algorithm must be stable. We get the desired sorting order by using \LFmapping: $j \mapsto \LF(j)$.

The range of outgoing edges $(u', v') \in E'$ from node $u' \in V'$ with lexicographic rank $i'$ is $[sp_{out}, ep_{out}] = [\bvOUT.\select(i', 1) + 1, \bvOUT.\select(i'+1, 1)]$. The edges are already sorted by keys $G'.\gkey(u')$. Because graph $G'$ is a path graph, we know that key $G'.\gkey(u')$ prefix-matches string $G'.\glabel(u') \cdot G'.\gkey(v')$. The sorting orders are therefore compatible. For every $j \in [sp_{in}, ep_{in}]$ for a node $v' \in V'$, having $\LF(j) \in [sp_{out}, ep_{out}]$ for a node $u' \in V'$ implies an edge $(u', v') \in E'$.

We use \emph{backward searching} for query $G'.\find(X)$. Let $X \in \patternset$ be a pattern. If $\abs{X} = 0$, query $G'.\find(X)$ returns the lexicographic range $[0, \abs{V'}-1]$ containing all nodes. Now assume that $\abs{X} \ge 1$ and that $G'.\find(X[i+1, \abs{X}-1]) = [sp_{i+1}, ep_{i+1}]$. We want to find the lexicographic range $G'.\find(X[i, \abs{X}-1])$, which is the union of sets $G'.\gpred(v', X[i])$ over nodes $v' \in G'.\find(X[i+1, \abs{X}-1])$. We map the node range $[sp_{i+1}, ep_{i+1}]$ to the range $[sp_{in}, ep_{in}]$ of incoming edges; the incoming edges to the corresponding range of outgoing edges $[sp_{out}, ep_{out}]$; and the outgoing edges to the range $[sp_{i}, ep_{i}] = G'.\find(X[i, \abs{X}-1])$:
\begin{align*}
[sp_{in}, ep_{in}] = [&\bvIN.\select(sp_{i+1}, 1) + 1, \\
 & \bvIN.\select(ep_{i+1}+1, 1)]; \\
[sp_{out}, ep_{out}] = [&\LF(sp_{in}, X[i]), \\
 & \LF(ep_{in}+1, X[i]) - 1]; \\
[sp_{i}, ep_{i}] = [&\bvOUT.\rank(sp_{out}, 1), \\
 & \bvOUT.\rank(ep_{out}, 1)].
\end{align*}
We can think this as a generalization of \LFmapping: $[sp_{i}, ep_{i}] = G'.\LF([sp_{i+1}, ep_{i+1}], X[i])$.

Query $G'.\locate(X)$ retrieves the values $G'.\gvalue(v')$ for nodes $v' \in G'.\find(X)$ and filters out duplicates. Instead of storing the values explicitly for all nodes, GCSA uses a \emph{sampling} scheme to save space. We assume that the nodes of the input graph $G = (V, E)$ are conveniently chosen integers. If $(u, v) \in E$ is the only outgoing edge from node $u$ and the only incoming edge to node $v$, it should be that $v = u+1$.

We sample the values $G'.\gvalue(v')$ for a node $v' \in V'$, (a) if there are multiple incoming edges to node $v'$; (b) if $v'$ is the source node $s$; or (c) if $G'.\gvalue(v') \ne \set{u+1 \mid u \in G'.\gvalue(u')}$ for the only incoming edge $(u', v') \in E'$. We may also sample the values for some nodes on long unary paths for performance reasons. If the set $G'.\gvalue(v')$ has not been sampled, we can derive it from sampled values by following the incoming edges.

If node $v' \in V'$ with lexicographic rank $i$ has only one predecessor, the lexicographic rank of the predecessor is $G'.\LF(i) = \bvOUT.\rank(\LF(\bvIN.\select(i, 1) + 1), 1)$. If the lexicographic rank $G'.\LF^{k}(i)$ corresponding to node $w' \in V'$ is the first sampled node we encounter, we know that $G'.\gvalue(v') = \set{w+k \mid w \in G'.\gvalue(w')}$.

Let $B_{S}[0, \abs{V'}-1]$ be a bitvector. If we have sampled the values for the node $v' \in V'$ with lexicographic rank $i$, we mark that as $B_{S}[i] = 1$. We can then determine the rank of node $v'$ among the sampled nodes as $j = B_{S}.\rank(i, 1)$. For each sampled node $v' \in V$, we store the size of the value set $\abs{G'.\gvalue(v')}$ in another bitvector $B_{V}$, using the same encoding as for bitvectors $\bvIN$ and $\bvOUT$. We store the samples in array $V_{S}$ in the same order, using $\log \abs{V}$ bits each. The sampled values for node $v'$ with rank $j$ among the sampled nodes can be found at $V_{S}[B_{V}.\select(j, 1) + 1, B_{V}.\select(j+1, 1)]$.

\section{Index Construction}\label{appendix:construction}

GCSA construction \cite{Siren2014} is based on the \emph{prefix-doubling} algorithm for suffix array construction \cite{Manber1993}. The original GCSA started from paths of length $1$ in the input graph, and then repeatedly \emph{joined} paths of length $k$ into paths of length $2k$, until each path had a distinct label. The resulting path graph was essentially an \orderk{\infty} pruned de~Bruijn graph and supported queries of any length.

We use a variant of that algorithm with GCSA2. Let $G = (V, E)$ be the input graph. We extract all paths of length $k$ (typically with $k = 16$) from graph $G$. For each path $P = v_{0} \dotsm v_{\abs{P}-1}$, we store several fields. \emph{Key} $P.\gkey$ encodes $G.\glabel(P)$ as a sequence of lexicographic ranks of \kmer{k}s. If $\abs{P}$ is not an integer multiple of $k$, the key consists of the \kmer{k} ranks for the lexicographically smallest \kmer{(\lceil \abs{P}/k \rceil \cdot k)} having $G.\glabel(P)$ as a prefix, followed by the rank of the last \kmer{k} in the largest such \kmer{(\lceil \abs{P}/k \rceil \cdot k)}. \emph{Value} $P.\gvalue$ is the start node $v_{0}$ of the path. We store the set of \emph{predecessor labels} $\set{c \in \Sigma \mid \abs{G.\gpred(v_{0}, c)} > 0}$ as $P.\gpred$. For each possible \emph{extension node} $v \in \set{v \in V \mid (v_{\abs{P}-1}, v) \in E}$, we create a separate copy of the path and store the node as $P.\gext = v$.

The construction uses several supporting structures. We build an \orderk{k} \emph{de~Bruijn graph} $G_{d} = (V_{d}, E_{d})$ of the path labels and encode it as a GCSA, using the predecessor labels $P.\gpred$ for determining the edges. Let $v_{0}, \dotsc, v_{\abs{V_{d}}-1}$ be the nodes of the de~Bruijn graph in lexicographic order by their keys. We use two additional arrays: the \emph{LCP array} $\LCP[0, \abs{V_{d}}-1]$, where $\LCP[i]$ is the length of the longest common prefix of keys $G_{d}.\gkey(v_{i-1})$ and $G_{d}.\gkey(v_{i})$ (with $\LCP[0] = 0$), and the \emph{last character array} $L[0, \abs{V_{d}}-1]$, where $L[i] = G_{d}.\gkey(v_{i})[k-1]$. The LCP array is stored as a wavelet tree for fast \emph{range minimum queries} \cite{Gagie2012a}.

Because we store path labels explicitly, we only do a limited number of \emph{doubling steps}, typically $2$ or $3$. After $d$ doubling steps, the length of the paths is $2^{d} k$, and we can use them to build a maximally pruned \orderk{(2^{d} k)} de~Bruijn graph. Each doubling step consists of a \emph{pruning} step, followed by an \emph{extension} step. The pruning step applies a limited form of Lemma~\ref{lemma:dbg-prune} to lexicographic ranges of paths. Given two paths $P$ and $P'$, we can determine the length of the longest common prefix of the path labels by using the keys $P.\gkey$ and $P'.\gkey$ and the LCP array. If all paths sharing a prefix start from the same node, we merge them into a single path $Q$ with $Q.\gkey$ based on the shared prefix and $Q.\gext = -1$.

The extension step transforms the current set of paths of length (up to) $k'$ into a set of paths of length (up to) $2k'$. If $P$ is a path with $P.\gext = -1$, we use it as such. If $P.\gext = P'.\gvalue$ for paths $P$ and $P'$, we create a new path $PP'$. We set $(PP').\gkey$ according to the concatenation of the path labels, take $\gvalue$ and $\gpred$ from path $P$, and take $\gext$ from path $P'$. If we have another path $Q$ with $Q.\gkey = P.\gkey$ such that $QP'$ is a path, and if $P'.\gext = -1$, all possible \kmer{2k'} extensions of label $G.\glabel(PP')$ are also labels of paths starting from node $Q.\gvalue$, and the other way around. Hence paths $PP'$ and $QP'$ can be represented by a single node in a pruned de~Bruijn graph.

The doubling steps are followed by the \emph{merging step}, which transforms the paths into the nodes of a maximally pruned de~Bruijn graph $G'' = (V'', E'')$. We merge the paths with identical keys into the nodes of a pruned de~Bruijn graph $G' = (V', E')$. If paths $P_{0}, \dotsm, P_{m-1}$ all have the same key, we create a node $v' \in V'$ with the shared key as $G'.\gkey(v')$ and with $G'.\gvalue(v') = \bigcup_{i=0}^{m-1} P_{i}.\gvalue$. We also store the union of predecessor labels as $v'.\gpred = \bigcup_{i=0}^{m-1} P_{i}.\gpred$. We then apply Lemma~\ref{lemma:dbg-prune} maximally, transforming graph $G'$ into graph $G''$.

Storing the paths and the graphs may require hundreds of gigabytes of memory when indexing whole-genome variation graphs. To avoid that, we keep them on \emph{disk} when possible. The subgraph corresponding to each chromosome is stored in a separate file, with paths sorted by their labels in lexicographic order. Extension steps are done separately for each of the chromosomes. The pruning step merges the sorted files. It keeps reading paths into a buffer, until it has found a maximal range of paths that can be merged. The merged path is written into the new file for that chromosome, and the original range of paths is removed from the buffer. The merging step works in a similar way.

After creating the nodes $V''$ of the maximally pruned \orderk{k'} de~Bruijn graph $G'' = (V'', E'')$, we \emph{build the index}. Sequences $\BWT$ and $\bvIN$ can be generated from the predecessor sets $v''.\gpred$, while the outdegree sequence $\bvOUT$ requires further processing. There is an edge $(u'', v'') \in E''$ if and only if string $c \cdot G''.\gkey(v'')$ prefix-matches key $G''.\gkey(u'')$ and $c \in v''.\gpred$. We determine the edges and produce the outdegree sequence by scanning the node file sequentially with $\sigma+1$ pointers. The pointer corresponding to node $v''$ scans the entire file, while each of the remaining $\sigma$ pointers scans only the range of nodes $u''$ with $G''.\glabel(u'') = c$ for a character $c \in \Sigma$. We also sample the nodes for $\locate$ queries during the same scans.

Checking whether key $G''.\gkey(u'')$ prefix-matches string $c \cdot G''.\gkey(v'')$ can be done by using the GCSA for de~Bruijn graph $G_{d} = (V_{d}, E_{d})$ and the last character array $L$. If the lexicographic rank of \kmer{k} $X$ is $i$, the lexicographic rank of \kmer{k} $(cX)[0, k-1]$ is $G_{d}.\LF(i, c)$. If \kmer{2k} $X$ is encoded with \kmer{k} ranks $(i, j)$, we can encode string $cX$ as a lexicographic range of \kmer{k} rank sequences, with $(G_{d}.\LF(i, c), G_{d}.\LF(j, L[i]), G_{d}.\LF(0, L[j]))$ as the lower bound and $G_{d}.\LF(\abs{V_{d}}-1, L[j])$ as the last rank in the upper bound. String $c \cdot G''.\gkey(v'')$ prefix-matches key $G''.\gkey(u'')$ if and only if the lexicographic ranges of the \kmer{k} rank sequences overlap.

\emph{Complex regions} of the variation graph must be pruned before indexing. While this happens before index construction begins, the construction algorithm has features that can make the pruning less destructive. Pruning heuristics often create regions that are completely missing from the index. The same mechanism that saves memory by having each chromosome in a separate file can be used to index \emph{overlapping subgraphs} without indexing any paths between them. By having the pruned graph in one file and the reference path in another file, we can guarantee that no region is completely missing from the index. We can also index \emph{selected paths} in complex regions by duplicating nodes for prefix-doubling and mapping the duplicates back to the original nodes during the merging step. If we index the paths corresponding to \emph{known haplotypes} in complex regions, we can guarantee that the index contains all observed variation.

\section{Simplified GCSA Encoding}\label{appendix:encoding}

Let $G' = (V', E')$ be a path graph with $G'.\gpred(v', c) \le 1$ for all nodes $v' \in V'$ and characters $c \in \Sigma$. This is true for de~Bruijn graphs, and also for maximally pruned de~Bruijn graphs (Lemmas~\ref{lemma:dbg-maximal} and \ref{lemma:pg-keys}). We can use the simplified encoding of the original GCSA \cite{Siren2014} with such path graphs.

We replace the sequences $\BWT$ and $\bvIN$ with \emph{indicator bitvectors} $B_{c}[0, \abs{V'}-1]$ for all $c \in \Sigma$. If node $v' \in V$ with lexicographic rank $i$ has a predecessor with label $c \in \Sigma$, we set $B_{c}[i] = 1$. The backward step becomes:
\begin{align*}
sp_{out} & = C[X[i]] + B_{X[i]}.\rank(sp_{i+1}, 1); \\
ep_{out} & = C[X[i]] + B_{X[i]}.\rank(ep_{i+1}+1, 1) - 1; \\
[sp_{i}, ep_{i}] & = [\bvOUT.\rank(sp_{out}, 1), \bvOUT.\rank(ep_{out}, 1)].
\end{align*}
Two expensive queries ($\bvIN.\select()$ and $\BWT.\rank()$) are replaced with a cheap $B_{c}.\rank()$.

Computing $G'.\LF(i)$ can expensive, as we have to look at $B_{c}[i]$ for all $c \in \Sigma$ to determine the character used in the backward step. If the alphabet is small, this is still faster than the $\select$ queries in the general encoding. We can further improve the time/space trade-off by compressing the bitvectors $B_{c}$ for rare characters (e.g.~$\baseN$, $\#$, and $\$$) and checking $B_{c}[i]$ first for the frequent characters (e.g.~bases) when computing $G'.\LF(i)$.

\section{Suffix Tree of a Path Graph}\label{appendix:extensions}

Let $G' = (V', E')$ be an \orderk{k} path graph, and let $v'_{0}, \dotsc, v'_{\abs{V'}-1}$ be its nodes in lexicographic order. The \emph{LCP array} of graph $G'$ is an array $\LCP[0, \abs{V'}-1]$, where $\LCP[i]$ is the length of the longest common prefix of keys $G'.\gkey(v'_{i-1})$ and $G'.\gkey(v'_{i})$, with $\LCP[0] = 0$. If we build a trie of keys $G'.\gkey(v')$ for $v' \in V'$ and compact the unary paths into single edges, the resulting tree can be considered the \emph{suffix tree} of graph $G'$.

The \emph{LCP interval tree} \cite{Abouelhoda2004} is an alternate representation of the suffix tree. Each node $v$ of the suffix tree is represented by the lexicographic range $[sp, ep]$ matching the path label $\ell(v)$ from the root to the node. These ranges can be determined from the LCP array. If we use the FM\nobreakdash-index with an LCP array supporting \emph{next/previous smaller value} queries and \emph{range minimum queries}, we can support the full functionality of the suffix tree \cite{Fischer2009a}.

If we build an $x$\nobreakdash-ary tree over the LCP array, with each internal node storing the minimum LCP value in the corresponding range, we can support the required queries in $\Oh{x \log_{x} \abs{V'}}$ time with $\Oh{\log_{x} \abs{V'}}$ random memory accesses, while using $\frac{x}{x-1} \abs{V'} \log k$ bits of space. Because the nodes of a path graph may match patterns that do not prefix-match their keys, we have to be careful with the suffix tree operations we use. By Lemma~\ref{lemma:pg-context}, the ranges returned by $\find$ queries always correspond to prefixes of the pattern. Hence we can safely use $\parent$ queries with such ranges.\footnote{The $\textsf{shorter}$ query in the variable-order de~Bruijn graph \cite{Boucher2014} is essentially a $\parent$ query.}

BWA-MEM \cite{Li2013} aligns reads to a reference genome by finding \emph{maximal exact matches} between the read and the reference. It indexes both the reference and its reverse complement, allowing it to extend the pattern in both directions. GCSA cannot use similar techniques, as we cannot guarantee that the length of the lexicographic range matching pattern $X$ is the same as the length of the range matching its reverse complement $\revcomp{X}$ (that the key set contains key $K$ if and only if it contains key $\revcomp{K}$). However, we can search for maximal exact matches by using \LFmapping{} and the $\parent$ operation \cite{Ohlebusch2010a}.

In an ordinary FM\nobreakdash-index, the length of a lexicographic range tells the number of distinct pointers in the range. In GCSA, each node $v' \in V'$ may have multiple values (pointers), and a value may occur in multiple nodes. As $\locate$ queries can be slow, we need another way to support efficient \emph{counting queries}. The problem is similar to determining the \emph{document frequency} of a pattern. Given a collection of documents, the document frequency of pattern $X$ is the number of distinct documents that contain occurrences of the pattern. We can use a bitvector of length $2n-d-1$, where $n$ is the total size of the document collection and $d$ is the number of documents, to compute the frequencies efficiently \cite{Sadakane2007a}.

Let $v$ be a suffix tree node corresponding to lexicographic range $[sp, ep]$, and let $\countq(v)$ be the document frequency of the label $\ell(v)$. If nodes $v_{0}, \dotsc, v_{m-1}$ are the children of node $v$, the number of \emph{redundant} documents in them is $R(v) = \sum_{i=0}^{m-1} \countq(v_{i}) - \countq(v)$. We create an array $R[0, n-2]$ based on the \emph{inorder traversal} of the suffix tree. If the $i$th internal node we encounter is our first visit to node $v$, we set $R[i] = R(v)$. We set $R[j] = 0$ for any subsequent visits to the same node. Range $R[sp, ep-1]$ covers the internal nodes in the subtree with node $v$ as the root. We can determine document frequencies as $\countq(v) = (ep+1-sp) - \sum R[sp, ep-1]$. If we encode array $R$ in unary, with value $x$ becoming $0^{x} 1$, we get a bitvector $B_{R}$, where we can compute sums $\sum_{i=a}^{b} R[i]$ with $\select$ queries as
$$
(B_{R}.\select(b+1, 1) - b) - (B_{R}.\select(a, 1) + 1 - a).
$$

For value counting, we use array $R$ to store the number of redundant values as above. We use another array $A[0, \abs{V'}-1]$ to store the number of additional values in each node $v'_{i} \in V'$ as $A[i] = \abs{G'.\gvalue(v'_{i})}-1$, and encode it as a bitvector $B_{A}$ in the same way as array $R$ above. The number of distinct values in range $[sp,ep] = G'.\find(X)$ for a pattern $X$ is
\begin{align*}
G'.\countq(X) & = G'.\countq(sp, ep) \\
 & = \sum_{i = sp}^{ep} (A[i] + 1) - \sum_{i = sp}^{ep - 1} R[i].
\end{align*}
The bitvectors are often highly compressible \cite{Gagie2015}, but GCSA already uses one of the compression schemes implicitly when it prunes the de~Bruijn graph.

\section{Using the Hypertext Index}\label{appendix:hypertext}

The hypertext index \cite{Thachuk2013} is based on graphs $G = (V, E)$, where the label $G.\glabel(v)$ of a node $v \in V$ is a string over alphabet $\Sigma$. The labels are indexed in FM\nobreakdash-index $F$, while the reverse labels are indexed in FM\nobreakdash-index $R$. When we search for a pattern $X \in \Sigma^{\ast}$ of length $\abs{X} \ge 2$, some of the matches may cross edges. In order to find matches crossing one edge, we search for suffixes $X[i, \abs{X}-1]$ in the forward index $F$ and the reverses of prefixes $X[0, i-1]$ in the reverse index $R$, for all $1 \le i \le \abs{X}-1$. For each value of $i$, we combine the partial matches into complete matches with a two-dimensional range query in the edge matrix $E$, using the lexicographic ranges for the reverse of $X[0, i-1]$ and for $X[i, \abs{X}-1]$ as the query ranges.

We often have to prune complex regions of the input graph before indexing it. This causes false negatives: paths that exist in the input graph but not in the index. We can avoid the false negatives with a generalization of the hypertext index. Instead of pruning the input graph heuristically, we create a primary graph based on known haplotypes and build a GCSA index for both strands of the graph. We then create a matrix of additional edges corresponding to potential \emph{recombinations} in the path graph, always crossing from the reverse complement strand to the forward strand. We search for pattern $X$ and its reverse complement $\revcomp{X}$ in the index, and combine the results $\find(X[i, \abs{X}-1])$ and $\find(\revcomp{X}[\abs{X}-i, \abs{X}-1])$ with a range query.

While graphs are a natural formalism for representing genetic variation, they cannot adequately represent certain types of \emph{rearrangements}. For example, if sequence $S$ can occur in different positions of the genome (e.g.~$ASBC$ and $ABSC$), we can either have a single copy or multiple copies of $S$ in the graph. Neither option is good in a reference genome. With a single copy, we create paths in the reference that do not correspond to any valid genome. With the second option, we lose the information that both copies of $S$ are the same sequence. One solution is to use \emph{context-free grammars}. As long as the grammar is non-nested, we can handle it with the hypertext index. We build a hypertext index for a high level graph, where each node is labeled with a nonterminal symbol, while each nonterminal expands into a subgraph indexed in GCSA.

\bibliographystyle{abbrv}
\bibliography{paper}

\end{document}